\documentclass[iop]{emulateapj}
\usepackage{natbib}
\usepackage{amsmath}
\usepackage{textcomp}
\usepackage[colorlinks=true,citecolor=blue]{hyperref}
\usepackage{graphicx,color}
\usepackage{times}
\usepackage{amssymb}

\newcommand{\lum}{erg\,s$^{-1}$}
\newcommand{\phflux}{\mbox{${\rm \, ph \,\, cm^{-2} \, s^{-1}}$}}
\newcommand{\ergflux}{\mbox{${\rm \, erg \,\, cm^{-2} \, s^{-1}}$}}
\newcommand{\fermi}{{\it Fermi}}

\slugcomment{ApJ accepted}

\shorttitle{The First GeV Flare of PKS 1502+036}
\shortauthors{Paliya and Stalin}

\begin{document}

\title{The First GeV Outburst of The Radio-loud Narrow Line Seyfert 1 Galaxy PKS 1502+036}

\author{Vaidehi S. Paliya$^{1,\,2}$ and C. S. Stalin$^1$} 
\affil{$^1$Indian Institute of Astrophysics, Block II, Koramangala, Bangalore-560034, India}
\affil{$^2$Department of Physics, University of Calicut, Malappuram-673635, India}
\email{vaidehi@iiap.res.in}

\begin{abstract}
The $\gamma$-ray loud narrow line Seyfert 1 ($\gamma$-NLSy1) galaxy PKS 1502+036 ($z=0.409$) exhibited its first $\gamma$-ray outburst on 2015 December 20. In the energy range of 0.1-300 GeV, the highest flux measured by {\it Fermi}-Large Area Telescope is (3.90 $\pm$ 1.52) $\times$ 10$^{-6}$ \phflux, which is the highest $\gamma$-ray flux ever detected from this object. The associated spectral shape is soft ($\Gamma_{0.1-300~{\rm GeV}}=2.57\pm0.17$) and this corresponds to an isotropic $\gamma$-ray luminosity of (1.2 $\pm$ 0.6) $\times$ 10$^{48}$ \lum. We generate the broadband spectral energy distribution (SED) during the GeV flare and reproduce it using a one zone leptonic emission model. The optical-UV spectrum can be explained by a combination of synchrotron and the accretion disk emission, whereas, the X-ray to $\gamma$-ray SED can be satisfactorily reproduced by inverse-Compton scattering of thermal photons originated from the torus. The derived SED parameters hint for the increase in the bulk Lorentz factor as a major cause of the flare and the location of the emission region is estimated as outside the broad line region but still inside torus. A comparison of the GeV flaring SED of PKS 1502+036 with that of two other $\gamma$-NLSy1 galaxies, namely, 1H 0323+342 ($z=0.061$) and PMN J0948+0022 ($z=0.585$), and also with FSRQ 3C 279 ($z=0.536$) has led to the conclusion that the GeV flaring SEDs of $\gamma$-NLSy1 galaxies resemble with FSRQs and a major fraction of their bolometric luminosity is emitted at $\gamma$-ray energies.
\end{abstract}


\section{Introduction}\label{sec:intro}
Blazars, a peculiar class of active galactic nuclei (AGN) with relativistic jets pointed towards the observer, are known to exhibit high amplitude $\gamma$-ray flux variations \citep[e.g.,][]{2011ApJ...733L..26A,2015ApJ...808L..48P}. Along with blazars, \fermi-Large Area Telescope \citep[\fermi-LAT;][]{2009ApJ...697.1071A} has also detected variable $\gamma$-ray emission from about half-a-dozen radio-loud narrow line Seyfert 1 (RL-NLSy1) galaxies \citep[e.g.,][]{2009ApJ...707L.142A,2011MNRAS.413.2365C,2015AJ....149...41P}. Though these sources host low luminosity jets compared to powerful flat spectrum radio quasars \citep[FSRQs;][]{2015A&A...575A..13F}, multiple episodes of $\gamma$-ray outbursts have been observed from some of the $\gamma$-ray emitting NLSy1 ($\gamma$-NLSy1) galaxies when their isotropic $\gamma$-ray luminosity exceeds 10$^{48}$ \lum~\citep[e.g.,][]{2011MNRAS.413.1671F,2012MNRAS.426..317D}. In general, FSRQs are known to emit such powerful GeV outbursts. Prior to this work, GeV flares have been observed only from 3 $\gamma$-NLSy1 galaxies, namely, 1H 0323+342 \citep[$z=0.061$;][]{2014ApJ...789..143P}, SBS 0846+513 \citep[$z=0.585$;][]{2012MNRAS.426..317D}, and PMN J0948+0022 \citep[$z=0.585$;][]{2011MNRAS.413.1671F,2015MNRAS.446.2456D}. A broadband study of these peculiar events are of great importance to understand the physical properties of relativistic jets in $\gamma$-NLSy1 galaxies at different black hole mass and accretion rate scales.

PKS 1502+036 ($z=0.409$) is one of the RL-NLSy1 galaxies detected in $\gamma$-ray band \citep{2009ApJ...707L.142A}. It is a faint but persistent $\gamma$-ray emitter \citep[e.g.,][]{2011MNRAS.413.2365C,2015AJ....149...41P} and subsequently included in the recently released third catalog of \fermi-LAT detected objects \citep[3FGL;][]{2015ApJS..218...23A}. It exhibits a compact core-jet structure \citep[][]{2012arXiv1205.0402O} and has a brightness temperature of $\sim$10$^{12}$ K \citep[][]{2008ApJ...685..801Y}. Rapid infra-red and intra-night optical variabilities of this source are also reported \citep[][]{2012ApJ...759L..31J,2013MNRAS.428.2450P}. Recently, this object was in a high $\gamma$-ray activity state when a $\gamma$-ray flux as high as $\sim$1 $\times$ 10$^{-6}$ \phflux, in 0.1$-$300 GeV energy range, was detected by LAT on 2015 December 20 \citep[][]{2015ATel.8447....1D}. This is the first GeV flare observed from this source. A quasi-simultaneous {\it Swift} telescope monitoring ensured the contemporaneous coverage of this peculiar event at lower energies as well \citep[][]{2015ATel.8450....1D}. Motivated by the availability of good quality data, we study this rare event following a variability and broadband spectral energy distribution (SED) modeling approach. Throughout, we adopt a $\Lambda$CDM cosmology with the Hubble constant $H_0=71$~km~s$^{-1}$~Mpc$^{-1}$, $\Omega_m = 0.27$, and $\Omega_\Lambda = 0.73$.
\section{Observations and Data Reductions}
\subsection{{\it Fermi}-Large Area Telescope Observations}\label{subsec:fermi}
We follow the standard data reduction procedure\footnote{http://fermi.gsfc.nasa.gov/ssc/data/analysis/documentation/} and describe it briefly. The recently released Pass 8 data, covering the period of the outburst (2015 December 16 to 2015 December 23 or MJD 57372$-$57379), are used to extract the 0.1$-$300 GeV SOURCE class events which are lying within 10$^{\circ}$ region of interest (ROI) centered at the 3FGL position of PKS 1502+036. To minimize the contamination from Earth limb $\gamma$-rays, we reject the events having zenith angle $>$90$^{\circ}$. The data analysis is performed with the ScienceTools (v10r0p5) package and post-launch instrument response function {\tt P8R2\_SOURCE\_V6}. The significance of $\gamma$-ray signal is determined by adopting a maximum likelihood (ML) test statistic TS=2$\Delta$log($\mathcal{L}$), where $\mathcal{L}$ represents the likelihood function between models with and without a point source at the position of source of interest \citep[][]{1996ApJ...461..396M}. All the sources present in the 3FGL catalog and lying within the ROI are considered and their spectral parameters are kept free to vary during the unbinned likelihood fitting. We also include the sources lying within 10$^{\circ}$ to 15$^{\circ}$ from the center of the ROI and their parameters are fixed to the 3FGL catalog values. We perform a first run of the ML analysis and the sources with TS$<$25 are removed from further analysis.

The $\gamma$-ray variability properties of the source are studied by generating light curves with various time binnings (1 day, 12 hr, 6 hr, and 3 hr). To generate the light curves, we freeze the photon indices of all the sources to the values obtained from the average analysis of the period of interest. Furthermore, to test for the presence/absence of a possible curvature, we apply various models to the $\gamma$-ray spectrum of the object. This includes a log-parabola ($N(E) = N_0(E/E_p)^{-\alpha-\beta~{\rm log}(E/E_p)}$, where $\alpha$ is the photon index at $E_p$, $\beta$ is the curvature index and $E_p$ is fixed at 300 MeV), and a power law model. We estimate 2$\sigma$ upperlimits for the time/energy bins with $\Delta F_{\gamma}/F_{\gamma} > 0.5$, where $\Delta F_{\gamma}$ is error in the flux $F_{\gamma}$, and/or 1$<$TS$<$9\footnote{TS = 9 corresponds to $\sim$3$\sigma$ detection \citep{1996ApJ...461..396M}.}. We do not consider the bins with TS$<$1 in the analysis. Statistical uncertainties are estimated at 1$\sigma$ level.

\subsection{{\it Swift} Observations}\label{subsec:swift}
A {\it Swift} target of opportunity observation was performed on 2015 December 22 \citep[][]{2015ATel.8450....1D}. {\it Swift} X-ray Telescope \citep[XRT;][]{2005SSRv..120..165B} observed the source in the most sensitive photon counting mode for a net exposure of $\sim$2.8 ksec. We perform the standard filtering and data analysis ({\tt xrtpipeline}) using HEASOFT (v 6.17) and the calibration database updated on 2015 November 5. To extract the source spectrum, we select a circular region of 30$^{\prime\prime}$, centered at the target. Background events are extracted from an annular region of inner and outer radii of 50$^{\prime\prime}$ and 150$^{\prime\prime}$, centered on the source, respectively. We combine the exposure maps using the task {\tt ximage} and {\tt xrtmkarf} is used to generate ancillary response files. The source spectrum is binned to have at least 1 count per bin. We adopt an absorbed power law \citep[$N_{\rm H}=3.93 \times 10^{20}$ cm$^{-2}$;][]{2005A&A...440..775K} and use C-statistics \citep[][]{1979ApJ...228..939C} to perform spectral fitting in XSPEC. The associated errors are calculated at 90\% confidence level.

{\it Swift} Ultraviolet/Optical Telescope \citep[UVOT;][]{2005SSRv..120...95R} observed PKS 1502+036 in all the six filters. We use the task {\tt uvotimsum} to add the individual frames. The source magnitudes are extracted using {\tt uvotsource}, corrected for galactic extinction following \citet[][]{2011ApJ...737..103S}, and converted to flux units using the zero points and conversion factors of \citet[][]{2011AIPC.1358..373B}.

\section{Results}\label{sec:results}
\subsection{Variability and Spectral Properties}\label{subsec:variability}
\begin{figure*}
\hbox{
      \includegraphics[width=\columnwidth]{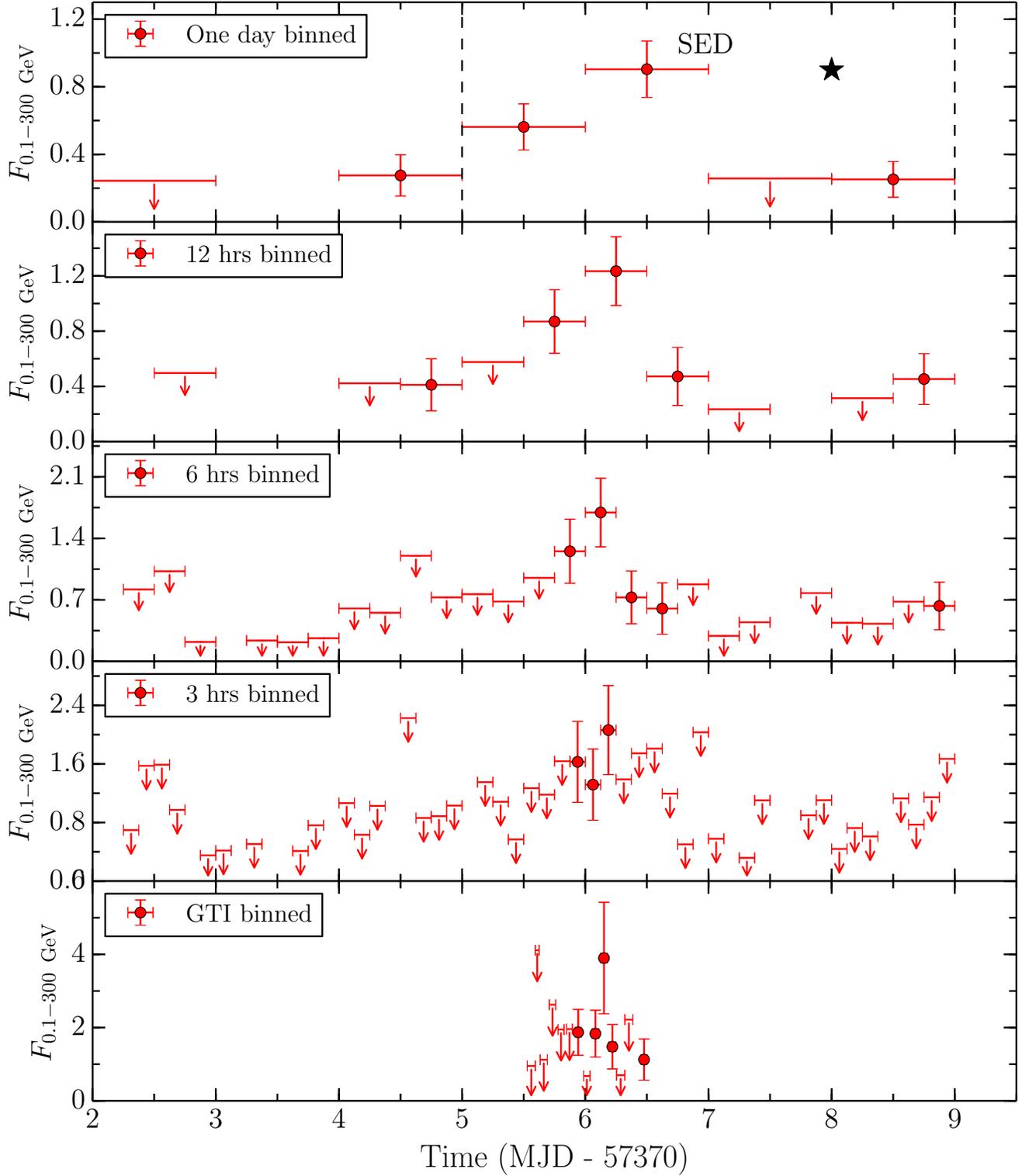}
     }
\caption{Gamma-ray flux variations of PKS 1502+036 covering the period of GeV outburst. Fluxes are in units of 10$^{-6}$ \phflux and downward arrows represnt the 2$\sigma$ upperlimits. In the top panel, the black star indicates the time of {\it Swift} monitoring and the dotted lines correspond to the period selected for SED generation and modeling.}\label{fig:fermi_lc}
\end{figure*}

The $\gamma$-ray flux variations of PKS 1502+036, covering the period of GeV outburst, are presented in Figure \ref{fig:fermi_lc}. As can be seen, the source entered in high activity state around MJD 57374. The rise in the flux appears smooth, as evident from the daily and 12 hr binned light curves and the maximum occurred on MJD 57376. Though the photon statistics is not good enough to perform a detailed flare profile fitting, the visual inspection of the 6 hr binned light curve indicates a slow rise and fast decay trend. Immediately after the flare, the source returned to low activity and was hardly detected after that, as evident from the 3 hr binned light curve. To determine the highest flux and also the shortest flux doubling/halving time, we generate the $\gamma$-ray light curve using bin sizes equal to good time intervals \citep[GTI, e.g.,][]{2011A&A...530A..77F}. A GTI is the shortest time period when LAT data can be considered `valid'\footnote{http://fermi.gsfc.nasa.gov/ssc/data/analysis/scitools/help/gtmktime.txt}. The maximum flux using this approach is derived as (3.90 $\pm$ 1.52) $\times$ 10$^{-6}$ \phflux~in the GTI bin 57376.1242$-$57376.1788, which is the highest $\gamma$-ray flux ever detected from PKS 1502+036 and is about 86 times its five year average value \citep[][]{2015AJ....149...41P}. No short term flux variability, of the order of hours or less, is detected.

FSRQs are known to exhibit a pronounced curvature in their $\gamma$-ray spectrum, especially during flaring episodes \citep[e.g.,][]{2015ApJ...808L..48P}. Such feature is also observed in the high activity state $\gamma$-ray spectrum of $\gamma$-NLSy1 galaxy SBS 0846+513 \citep[][]{2015AJ....149...41P}. With this in mind, we search for the presence of spectral curvature in the $\gamma$-ray spectrum of PKS 1502+036 by deriving the TS of the curvature \citep[TS$_{\rm curve}$ = 2(log $\mathcal{L}$(log-parabola) $-$ log $\mathcal{L}$(PL));][]{2012ApJS..199...31N}. We obtained a TS$_{\rm curve}=2.01$, indicating the absence of the curvature. Furthermore, the photon index obtained from average analysis of the period of interest is 2.57 $\pm$ 0.17 which is similar to that obtained from its five year average value \citep[][]{2015AJ....149...41P}.

During the flare, the 0.3$-$10 keV X-ray flux increases by a factor of $\sim$1.5 with respect to the low activity state (3.52$^{+0.73}_{-0.62}\times$ 10$^{-13}$ \ergflux, see Table \ref{tab:sed_flux1}) studied in our earlier work \citep[][]{2013ApJ...768...52P}. There are hints for the spectral hardening ($\Gamma_{0.3-10~{\rm keV}}=1.33^{+0.56}_{-0.55}$), however a strong claim cannot be made due to large errors. Comparing to the same low activity state, the source brightened by $\sim$0.5$-$0.7 magnitudes in optical-UV bands during the GeV flare, as revealed by {\it Swift}-UVOT monitoring (Table \ref{tab:sed_flux1}).

\subsection{Spectral Energy Distribution}
The broadband SED of PKS 1502+036 is generated for the period MJD 57375$-$57379 (see Figure \ref{fig:fermi_lc}). This period is chosen on the basis of the availability of multi-frequency data and also the requirement to generate good quality LAT spectrum. The generated SED is shown in the top panel of Figure \ref{fig:SED} and the associated flux values are reported in Table \ref{tab:sed_flux1}. For a comparison, we also show a relatively low activity state SED considered in our earlier work \citep[][]{2013ApJ...768...52P}.

\begin{figure*}
      \includegraphics[width=\columnwidth]{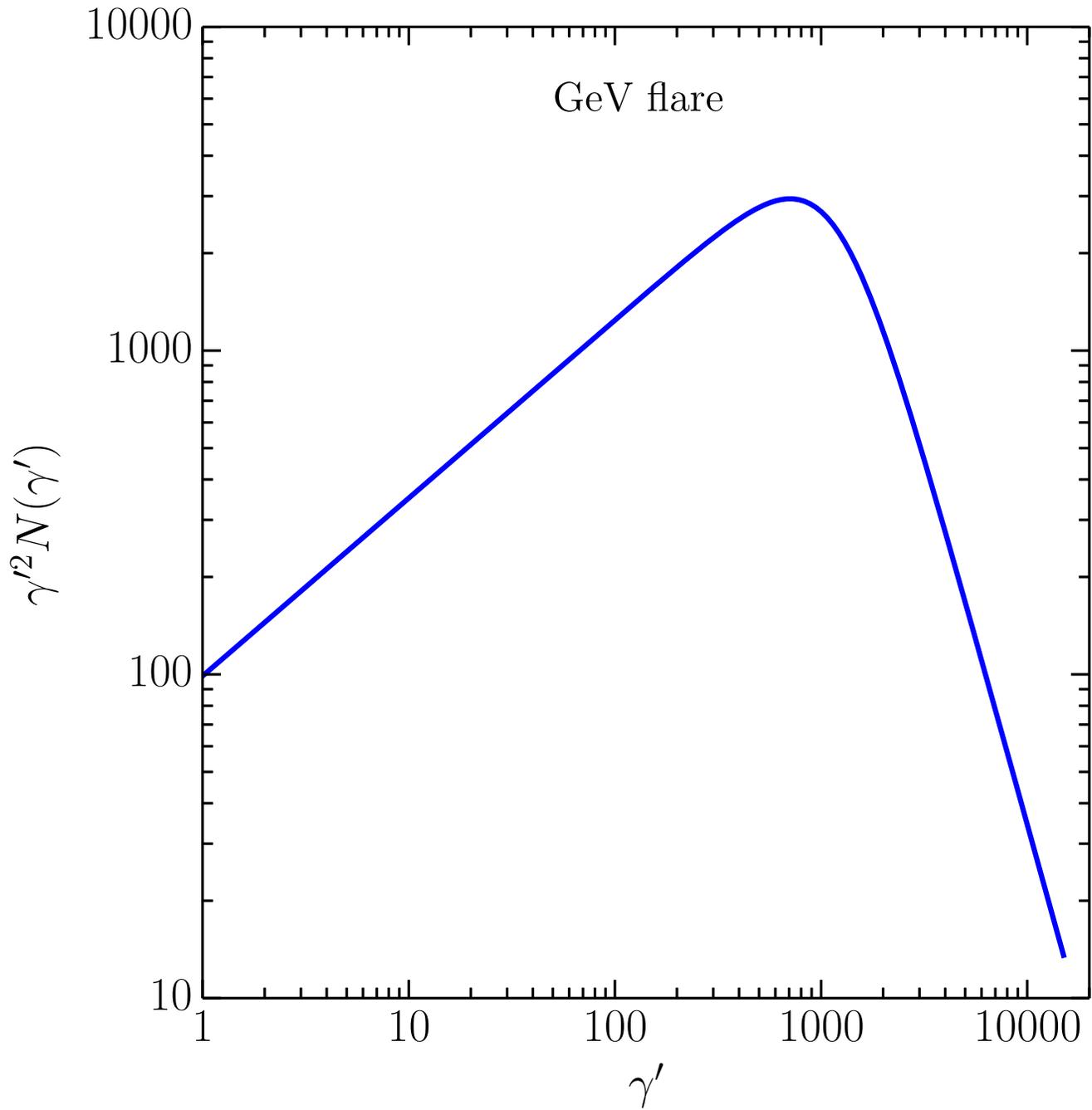}
\caption{Electron energy distribution used to model the GeV flaring SED of PKS 1502+036.}\label{fig:EED}
\end{figure*}

The generated broadband SED is modeled following the prescriptions of \citet{2009MNRAS.397..985G} and here it is briefly described. The emission region is assumed to be spherical and moves with a bulk Lorentz factor $\Gamma$. The relativistic electrons present in the emission region emit via synchrotron and inverse Compton processes and are assumed to follow a smooth broken power law energy distribution (Figure \ref{fig:EED})
\begin{equation}
 N'(\gamma')  \, = \, N_0\, { (\gamma'_{\rm b})^{-p} \over
(\gamma'/\gamma'_{\rm b})^{p} + (\gamma'/\gamma'_{\rm b})^{q}},
\end{equation}
where $p$ and $q$ are the energy indices before and after the break energy ($\gamma'_{\rm b}$), respectively, and prime quantities are in comoving frame. The spectrum of the accretion disk is considered as a multi-temperature blackbody \citep[][]{2002apa..book.....F}. Above and below the disk, we assume the presence of X-ray corona whose spectrum is considered as a cut-off power law. The broad line region (BLR) and the torus are assumed as spherical shells located at distances $R_{\rm BLR}=10^{17}L^{1/2}_{\rm disk,45}$ cm and $R_{\rm torus}=10^{18}L^{1/2}_{\rm disk,45}$ cm, respectively, where $L_{\rm disk,45}$ is the accretion disk luminosity in units of 10$^{45}$ \lum \citep[e.g.,][]{2007ApJ...659..997K,2009ApJ...697..160B}. We calculate the relative contributions of these components with respect to the distance from the central black hole following \citet{2009MNRAS.397..985G}. We derive the black hole mass and the accretion disk luminosity as 10$^{7.65}$ $M_{\odot}$ and 10$^{44.78}$ \lum, respectively, by reproducing the low activity state optical-UV spectrum with a standard optically thick, geometrically thin \citet{1973A&A....24..337S} accretion disk model (see Figure \ref{fig:SED}). The accretion disk luminosity can be constrained from the observations, provided the big blue bump is visible and assuming a fix accretion efficiency (considered as $\eta_{\rm disk}=10$\%). This leaves the black hole mass as the only free parameter. A large black hole mass implies a larger accretion disk surface and that, in turn, hints the lower value of the peak disk temperature needed to emit a fixed accretion disk luminosity \citep[see,][]{2002apa..book.....F}, thus implying a `redder' spectrum. The hole hole mass and the accretion disk luminosity derived by us (4.5 $\times$ 10$^{7} M_{\odot}$ and 6 $\times$ 10$^{44}$ \lum, respectively) are similar to that reported by \citet{2014Natur.515..376G}. It should be noted that the black hole mass derived from this method agrees, on average, within a factor of $\sim$4 to that obtained from virial relationship \citep[][]{2015MNRAS.448.1060G}. On the other hand, a broader limiting range of the disk luminosity can be set by ensuring 10$^{-2}L_{\rm Edd}<L_{\rm disk}<L_{\rm Edd}$. The lower limit assumes the accretion disk is radiatively efficient and the upper limit ensures the source to be sub-Eddington. Furthermore, along with synchrotron photons, electrons also scatter thermal photons entering from the accretion disk, the BLR, and the dusty torus via external Compton (EC) process. Various jet powers are derived following the prescriptions of \citet{2008MNRAS.385..283C}. In particular, kinetic jet power is calculated by assuming the protons to be cold and having an equal number density to that of electrons. The viewing angle is assumed as $\theta_{\rm view}=3^{\circ}$, a value typically considered for blazars \citep[e.g.,][]{2015MNRAS.448.1060G}. We model both GeV flaring and low activity SEDs using the methodology described above and show them in Figure \ref{fig:SED}. The associated modeling parameters are given in Table \ref{tab:sed_par}.
 
\section{Discussion}\label{sec:dscsn}
At the peak of the flare, the highest $\gamma$-ray flux measured is (3.90 $\pm$ 1.52) $\times$ 10$^{-6}$ \phflux~and the associated photon index is 2.57$\pm$0.17. This corresponds to an isotropic $\gamma$-ray luminosity ($L_{\gamma}$) of (1.22 $\pm$ 0.57) $\times$ 10$^{48}$ \lum, which is $\sim$96 times larger than its five year average value \citep[][]{2015AJ....149...41P}. Furthermore, the $\gamma$-ray luminosity in the jet frame would be $L_{\gamma, em} \simeq L_{\gamma}/2\Gamma^{2} \simeq$ 1.7 $\times$ 10$^{45}$ \lum, assuming $\Gamma=19$, obtained from SED modeling. Interestingly, this is a significant fraction of the total available accretion power ($\sim$35\%, $L_{\rm acc} \simeq L_{\rm disk}/\eta_{\rm disk} \simeq 6 \times 10^{45}$ erg s$^{-1}$; assuming radiative efficiency $\eta_{\rm disk}$ = 10\%) and also comparable to Eddington luminosity ($\sim$33\%, $L_{\rm Edd}\approx$ 5.7 $\times$ 10$^{45}$ \lum).

A careful examination of the jet energetics reveals few interesting features. First, the power spent by the jet in the form of radiation is larger than the sum of the jet power in electrons and magnetic field, during the GeV flare. This indicates the requirement for another power source to account for the radiative power and the assumption of the cold protons present in the jet would be the most plausible option. It should be noted that this feature has already been seen in the GeV flares of other $\gamma$-NLSy1 galaxies (e.g., Paliya et al. 2016, ApJ, in press), in addition to previously known luminous FSRQs and radio galaxies \citep[][]{2014Natur.515..376G,2015ApJ...799L..18T}. Furthermore, considering the radiative efficiency of the jet as $\eta_{\rm rad}=P_{\rm r}/P_{\rm jet}$, we find $\eta_{\rm rad}=0.32$ and 0.04 during the GeV flare and the low activity state, respectively. This suggests an efficient conversion of the jet kinetic power to the radiative power during the $\gamma$-ray flaring activity.

\begin{figure*}
\hbox{\hspace{-0.5cm}
      \includegraphics[width=8cm]{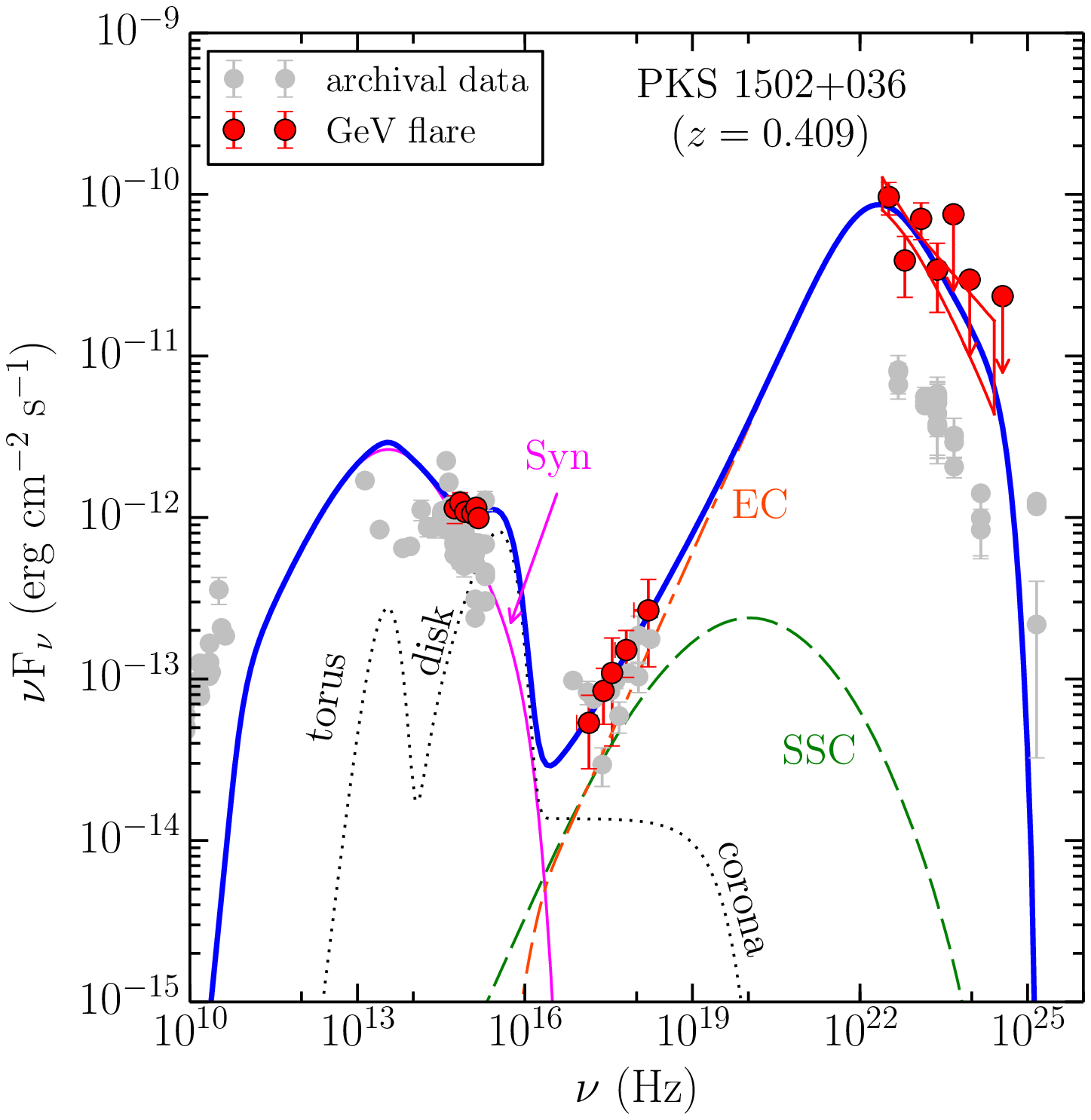}
      \includegraphics[width=8cm]{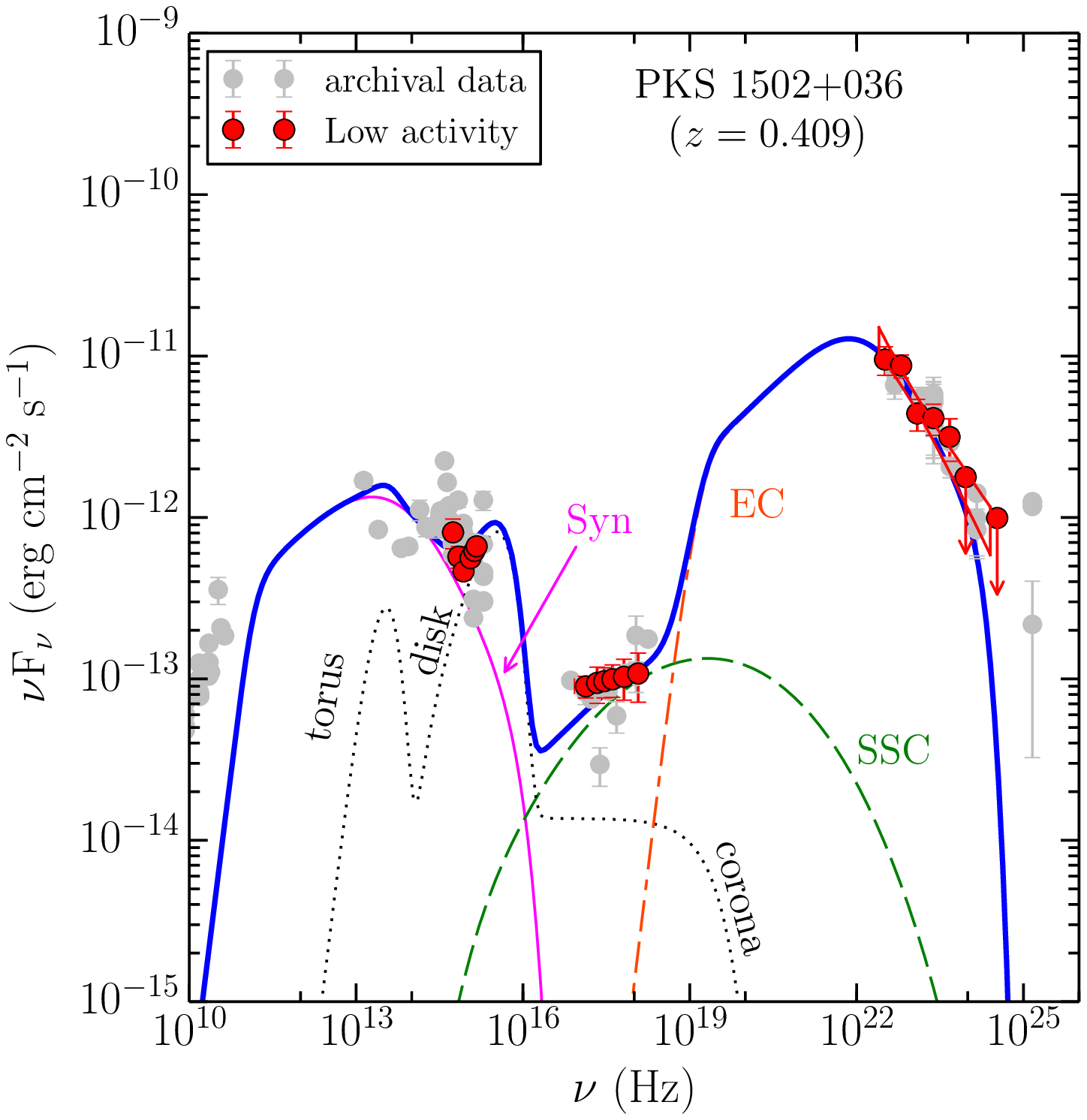}
     }
\hbox{\hspace{-0.5cm}
      \includegraphics[width=8cm]{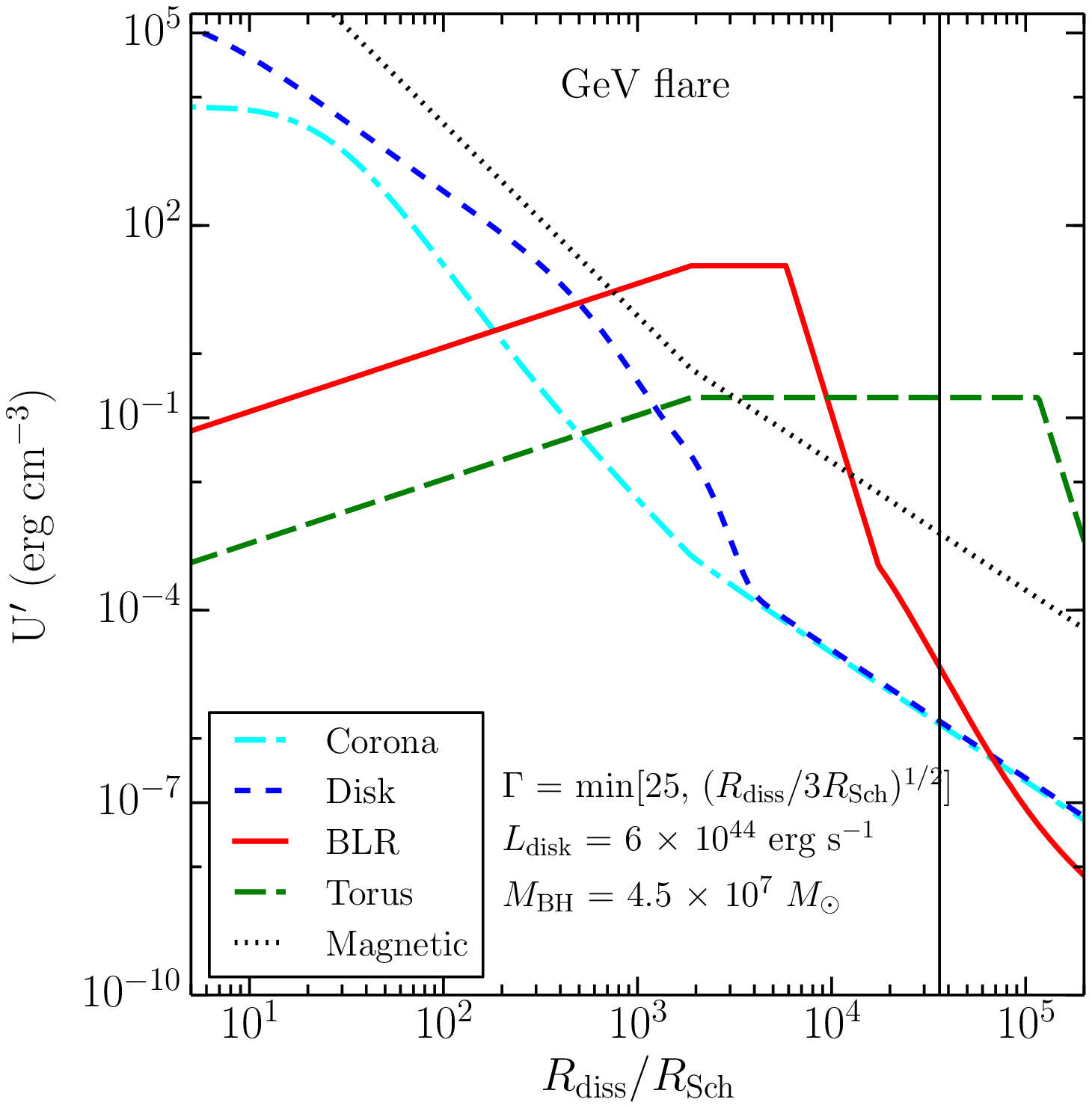}
      \includegraphics[width=8cm]{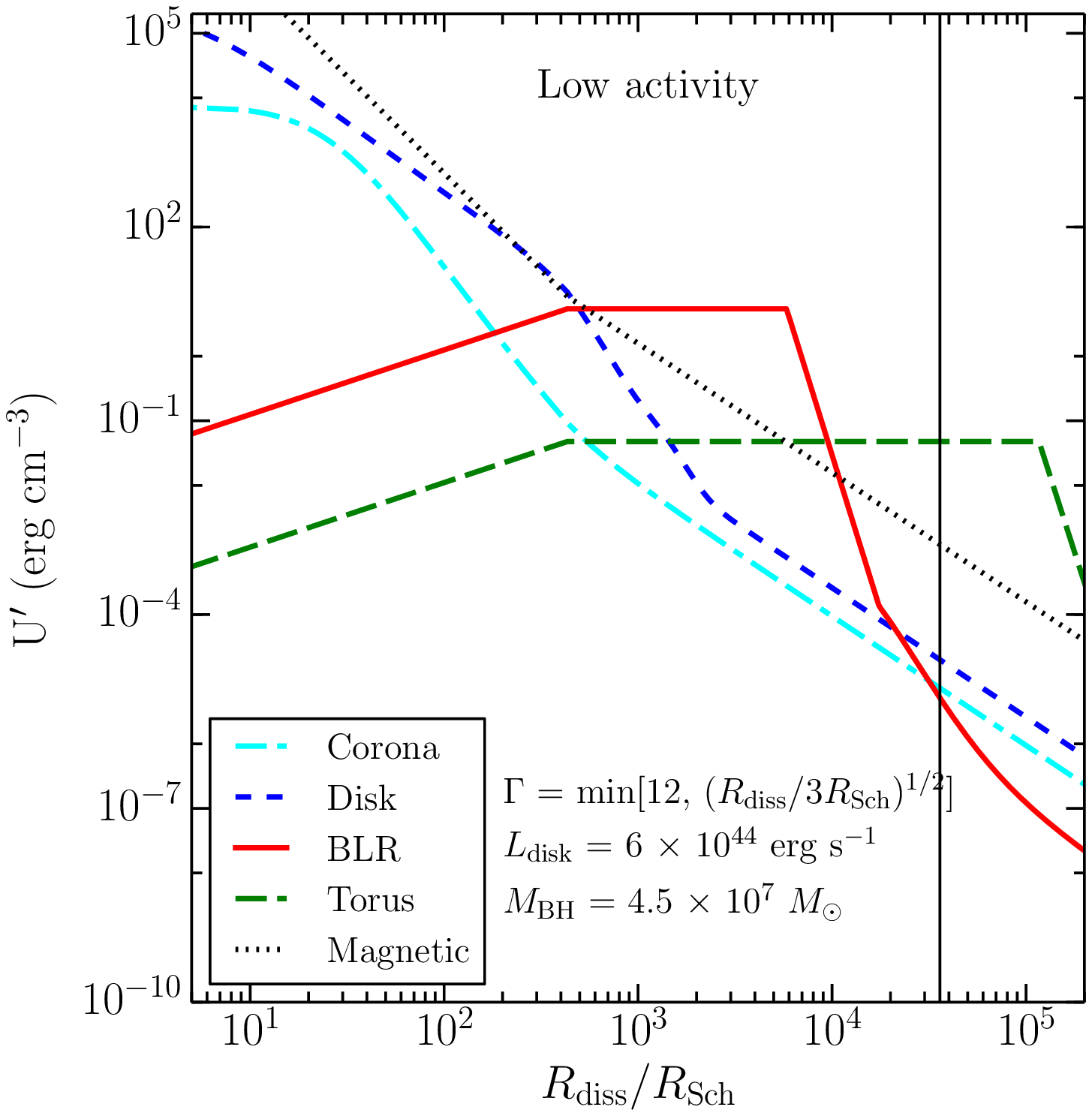}
     }
\caption{Top: The modeled SEDs of PKS 1502+036 during the period of GeV outburst (left) and a low activity state (right). Red circles denote the quasi-simultaneous observations, whereas, grey circles refer to the archival data. Pink thin solid, green long dashed, and orange dash-dash-dot lines represent synchrotron, SSC, and EC processes. Blue thick solid line is the sum of the out output of all the radiative mechanisms. Bottom: Variation of the comoving frame radiation energy densities as a function of the distance from the central black hole. The vertical line refers to the location of the emission region.}\label{fig:SED}
\end{figure*}

The high activity state optical-UV spectrum of PKS 1502+036 can be explained as a combination of synchrotron and the accretion disk radiation. Compared to the low activity state where a break is observed (that can be interpreted as a falling synchrotron and a rising accretion disk radiation), the shape of the high state optical-UV SED hints for enhanced synchrotron emission during the flare. Though the X-ray flux levels appear similar in both the activity states, the X-ray spectrum becomes harder during the GeV flare, which we interpret as a rising EC process. The $\gamma$-ray window of the SED can be well explained by EC mechanism with seed photons provided by IR-torus, similar to that reported by \citet[][]{2013ApJ...768...52P}. This sets the location of the emission region far out from the BLR but still inside the dusty torus (Table \ref{tab:sed_par}). The location of the emission region is derived from the following two constraints. First, we did not find any short timescale flux variability in the $\gamma$-ray band. Assuming it to be  $\sim$1 day \citep[similar to that taken by][]{2013ApJ...768...52P} and considering the conical geometry of the jet with semi opening angle $\theta_{\rm jet}=\theta_{\rm view}=3^{\circ}$ ($\sim 1/\Gamma$), we have
\begin{equation}
R_{\rm diss} = 20R_{\rm blob} \lesssim 20 \frac{ct_{\rm var}\delta}{(1+z)} \approx 0.16 {\rm pc}
\end{equation}
where $t_{\rm var}$ is the timescale of variability and $\delta=1/[\Gamma(1-\beta \cos\theta_{\rm view})]$. Now, the size of the BLR and the dusty torus are derived as 0.02 parsec and 0.5 parsec, respectively, thus indicating the emission region to be located outside the BLR but inside the torus. Second, in one zone models, the location of the synchrotron peak constrains the position of the IC peak. Since, the characteristic frequency of BLR photons ($\sim$10$^{15}$ Hz) is higher than that of torus photons ($\sim$10$^{13}$ Hz), corresponding EC-BLR will peak at higher frequencies than EC-torus process. In other words \citep[see,][]{2007ApJ...665..980T}
\begin{equation}
\nu_{\rm peak, obs} \simeq \frac{\nu_{\rm seed}\Gamma^2\gamma_{\rm b}^{\prime 2}}{(1+z)}
\end{equation}
where $\nu_{\rm peak, obs}$ is the observed peak frequency of the EC process, $\nu_{\rm seed}$ is the characteristic frequency of seed photons for EC mechanism, and $\gamma'_{\rm b}$ is the break Lorentz factor constrained from the location of the synchrotron peak (Table \ref{tab:sed_par}). This gives, $\nu_{\rm peak}\approx 10^{24}$ Hz or $10^{22}$ Hz, provided the seed photons are originating from the BLR or the dusty torus. It is clear that EC-BLR process cannot explain the observed soft $\gamma$-ray spectrum that demands the IC peak to lie at lower frequencies. This indicates the EC-torus as a plausible mechanism to reproduce the observed $\gamma$-ray spectrum. Furthermore, we can neglect EC-BLR emission by assuming the emission region to be sufficiently far out from the BLR where its contribution to comoving frame total radiation energy density is negligible. In the bottom panel of Figure \ref{fig:SED}, we plot the variation of the radiation energy densities as a function of the distance from the central engine and as can be seen, the emission region is probably located outside the BLR.

The $\gamma$-ray spectral shape remains similar in both the activity states, however, the flux increases by an order of magnitude. Comparing the SED parameters obtained during the flare with that derived during the low activity state, we find increase in the bulk Lorentz factor as a major cause of the outburst. This is also supported by the fact that though optical-UV and X-ray fluxes also increase, a major enhancement is seen in the $\gamma$-ray band only. In the emission region frame, the synchrotron emissivity can be adopted as \citep[e.g.,][]{1991pav..book.....S}
\begin{equation}
\label{eq:appaemiss}
j'_{\rm syn}(\nu')\approx\frac{\sigma_{\rm T}cB^2}{48 \pi^2}\nu_{\rm L}^{-\frac{3}{2}}
  N'\left(\sqrt{\frac{\nu'}{\nu_{\rm L}}}\right)\nu^{\prime{\frac{1}{2}}}
\end{equation}
where $\nu_{\rm L}$ is the  Larmor frequency. The EC emissivity can be taken as \citep{2012MNRAS.419.1660S}
\begin{equation}
\label{eq:ecemiss}
j'_{\rm EC}(\nu')\approx\frac{c\sigma_{\rm T}U^\star}{8\pi\nu^\star}
\left(\frac{\Gamma \nu'}{\nu^\star}\right)^{\frac{1}{2}} 
N'\left[\left(\frac{\nu'}{\Gamma\nu^\star}\right)^{\frac{1}{2}}\right]
\end{equation}
where $U$ is the external photon density and starred quantities are in the AGN frame. Comparing Equation~\ref{eq:appaemiss} and \ref{eq:ecemiss} we find that the excess in EC emissivity can be achieved by enhancing the bulk Lorentz factor of the jet, without altering the synchrotron emissivity. There are few other factors that changes such as the increase in the energy of the injected electrons and a slight enhancement of the magnetic field (Table \ref{tab:sed_par}). Overall, these parameters and/or their combination can reproduce the observed brightening seen in the $\gamma$-ray band. Furthermore, though we explain both X-ray and $\gamma$-ray emission via same EC mechanism, the flux increment appears relatively lower in the former. This is primarily because we reproduce low activity state X-rays via synchrotron self Compton (SSC) process and it also has some contribution during the GeV flare (see Figure \ref{fig:SED}). Similar to optical-UV, following Equation~\ref{eq:appaemiss} and \ref{eq:ecemiss}, a relatively lower variability at X-rays with respect to $\gamma$-rays can be understood. Furthermore, the value of the minimum energy of electrons ($\gamma'_{\rm min}$) also changes between the two states, which is due to different spectral shape observed at X-ray energies. A soft X-ray spectrum hints to be originated either from X-ray corona \citep[][]{2014ApJ...789..143P} or from SSC process. This suggests EC process to contribute negligibly in the X-ray band, requiring a relatively large $\gamma'_{\rm min}$. On the other hand, during the flare, the X-ray spectrum becomes flatter and to explain it via EC mechanism, we need $\gamma'_{\rm min}=1$.

\begin{figure*}
      \includegraphics[width=\columnwidth]{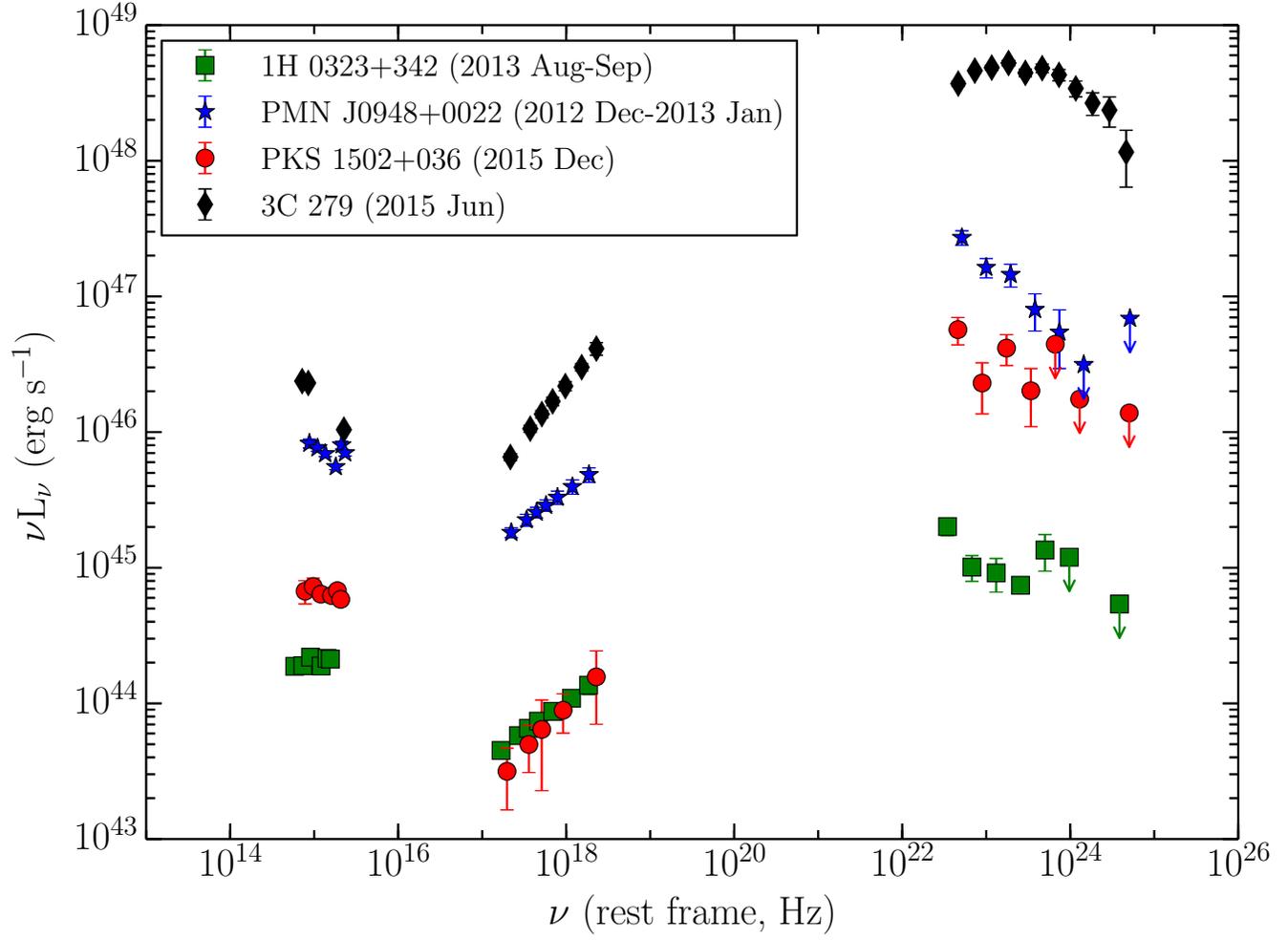}
\caption{Broadband SEDs of 3 $\gamma$-NLSy1 galaxies covering the period of their GeV outbursts. For a comparison, we also show the flaring state SED of the FSRQ 3C 279.}\label{fig:SED_comp}
\end{figure*}

It is of great interest to compare the flaring state SED of PKS 1502+036 with that of other GeV flaring $\gamma$-NLSy1 galaxies. With this in mind, we generate SEDs of two other $\gamma$-NLSy1 galaxies covering their GeV outbursts, namely, 1H 0323+342 \citep[][]{2014ApJ...789..143P} and PMN J0948+0022 \citep[][]{2015MNRAS.446.2456D}. We do not include another GeV flaring $\gamma$-NLSy1 galaxy SBS 0846+513 because there were no simultaneous multi-wavelength observations at the time of its GeV flare \citep[see,][]{2012MNRAS.426..317D}. In luminosity vs. frequency plane, we plot all the SEDs and the results are presented in Figure \ref{fig:SED_comp}. In this plot, PKS 1502+036 is slightly more luminous at optical-UV energies than 1H 0323+342, however, both the shape and the luminosity of the X-ray spectrum of PKS 1502+036 are similar to that of 1H 0323+342. At $\gamma$-ray energies, PKS 1502+036 is more luminous than 1H 0323+342. The apparent differences in the $\gamma$-ray band can be understand in terms of the higher Doppler boosting in the case of PKS 1502+036. For 1H 0323+342, \citet[][]{2014ApJ...789..143P} noticed the bulk Lorentz factor of the flaring emission region as $\Gamma=8$, whereas for PKS 1502+036, it is 25. It should be noted that $\gamma$-ray luminosity also depends on the external photon energy density which itself depends on the bulk Lorentz factor. In particular, assuming the emission region to lie inside BLR, we have \citep[e.g.,][]{2009MNRAS.397..985G}
\begin{equation}
U'_{\rm BLR} \sim \Gamma^2 \frac{\eta_{\rm BLR}L_{\rm disk}}{4\pi R^2_{\rm BLR}c} = \frac{\Gamma^2}{12\pi},
\end{equation}
assuming $R_{\rm BLR}\sim10^{17}L^{1/2}_{\rm d,45}$ cm and $\eta_{\rm BLR}$ is the fraction of the accretion disk luminosity reprocessed by BLR and here taken as 10\%. A similar relation holds for the torus energy density also. Both the BLR and the torus adjust their sizes according to the accretion disk luminosity so as to give a constant radiative energy density in the lab frame. Furthermore, there are few other factors which one should take into account, e.g., the location of the emission region, which was found to be inside the BLR for 1H 0323+342 and outside the BLR for PKS 1502+036. This suggests a higher magnetic field for the case of former, implying a high level of synchrotron and SSC emission. A high synchrotron emission is not reflected in the optical-UV spectrum of 1H 0323+342 due to the fact that its synchrotron radiation peaks at around sub-milimeter frequencies and thus the observed optical-UV radiation is dominated by the disk emission, even during the GeV flare \citep[see,][]{2014ApJ...789..143P}. On the other hand, modeling of the flaring SED of PKS 1502+036 indicates a significant contribution of the synchrotron mechanism at optical-UV energies. This could be one of the reason for the higher optical-UV luminosity of PKS 1502+036. A possible explanation for the difference in the locations of the synchrotron peaks for 1H 0323+342 and PKS 1502+036 can be given on the basis of the difference in their accretion disk luminosities. The accretion disk of 1H 0323+342 is more powerful than that of PKS 1502+036, though it is not reflected in the observed optical-UV band. This is because the peak of the disk luminosity occurs at unobserved far-UV energies for 1H 0323+342 \citep[][]{2009ApJ...707L.142A,2014ApJ...789..143P,2014PASJ...66..108I}. A higher disk luminosity means a denser external photon field which, in turn, indicates a faster cooling of the emitting electrons before they could reach to higher energies. Accordingly, synchrotron peak will lie at lower frequencies and may not contribute significantly in the optical-UV band. Furthermore, a high SSC emission and also contribution from the EC scattering of the accretion disk photons \citep[EC-disk;][]{2014ApJ...789..143P} can explain the similar X-ray luminosity of 1H 0323+342 with PKS 1502+036. A comparison of the SED of PKS 1502+036 with PMN J0948+0022 reveals that the latter one is more powerful at all the energies. The accretion disk luminosity of PMN J0948+0022 is derived as 9 $\times$ 10$^{45}$ \lum~\citep[][]{2012A&A...548A.106F} and 1.18 $\times$ 10$^{46}$ \lum~\citep[][]{2015A&A...575A..13F} based on the accretion disk model fitting and optical spectroscopic approach, respectively. This suggests the accretion disk of PMN J0948+0022 to be more luminous than PKS 1502+036 and can be clearly seen in the optical-UV band. The higher luminosity of the former at X-ray and $\gamma$-ray energies could be due to stronger boosting and larger power of injected electrons \citep[][]{2015MNRAS.446.2456D}.

In our earlier work, we compared the low activity SEDs of $\gamma$-NLSy1 galaxies with FSRQ 3C 454.3 ($z=0.859$) and BL Lac object Mrk 421 \citep[$z=0.031$;][]{2013ApJ...768...52P} and found them to resemble more to FSRQs. Later, few other observations, such as curved $\gamma$-ray spectrum, also supported this finding \citep[e.g.,][]{2015AJ....149...41P}. Therefore, it is interesting to test whether GeV flaring SEDs of $\gamma$-NLSy1 galaxies are also similar to that of FSRQs. We, therefore, generate SED of the FSRQ 3C 279 covering its recent GeV outburst in 2015 June \citep[][]{2015ApJ...808L..48P} and plot it along with other $\gamma$-NLSy1 galaxies in Figure \ref{fig:SED_comp}. As can be seen, 3C 279 is more luminous than any $\gamma$-NLSy1 galaxy, especially at $\gamma$-ray energies. The overall shapes of the SEDs of 3C 279 and $\gamma$-NLSy1 galaxies are similar, thus suggesting similar mechanisms to be working for the observed flaring behaviors, with more extreme parameters for 3C 279. Finally, the Compton dominance (the ratio of the inverse-Compton to synchrotron peak luminosities) of all the GeV flaring $\gamma$-NLSy1 galaxies is found to be greater than unity, a feature generally exhibited by FSRQs like 3C 279. Therefore, it can be concluded that the flaring state behavior of $\gamma$-NLSy1 galaxies resembles more to powerful FSRQs.

\acknowledgments
We are grateful to an anonymous referee for constructive suggestions that helped to improve the manuscript. This research has made use of data, software and/or web tools obtained from NASA’s High Energy Astrophysics Science Archive Research Center (HEASARC), a service of Goddard Space Flight Center and the Smithsonian Astrophysical Observatory. This research has made use of the XRT Data Analysis Software (XRTDAS) developed under the responsibility of the ASDC, Italy. Data from the Steward Observatory spectropolarimetric monitoring project were used. This program is supported by Fermi Guest Investigator grants NNX08AW56G, NNX09AU10G, and NNX12AO93G. Use of {\it Hydra} cluster at Indian Institute of Astrophysics is acknowledged.
\bibliographystyle{apj}
\bibliography{Master}

\newpage
\begin{table}
\begin{center}
{
\caption{Summary of SED generation analysis. {\it Fermi}-LAT, {\it Swift}-XRT, and {\it Swift}-UVOT fluxes are in units of 10$^{-7}$ \phflux, 10$^{-13}$ \ergflux, and 10$^{-12}$ \ergflux, respectively.}\label{tab:sed_flux1}
\begin{tabular}{lcccccc}
\hline\hline
 & & & {\it Fermi}-LAT & &  & \\
 Activity & & Flux$_{0.1-300~{\rm GeV}}$ & $\Gamma_{0.1-300~{\rm GeV}}$ & Test Statistic & \\
 \hline
 GeV flare    & & 4.41 $\pm$ 0.61 & 2.57 $\pm$ 0.16 & 109.69 & & \\
 Low activity & & 0.51 $\pm$ 0.07 & 2.58 $\pm$ 0.10 & 172.12 & & \\
  \hline
 &  &  & {\it Swift}-XRT  & &  & \\
 Activity & & Flux$_{0.3-10~{\rm keV}}$ & $\Gamma_{0.3-10~{\rm keV}}$ & C-Statistics & &\\
\hline
 GeV flare    & & 5.26$^{+3.08}_{-1.96}$ & 1.33$^{+0.56}_{-0.55}$ & 15.48/23 & & \\
 Low activity & & 3.52$^{+0.73}_{-0.62}$ & 1.92$^{+0.27}_{-0.27}$ & 66.20/82 & & \\
 \hline
 & & & {\it Swift}-UVOT & & &  \\
 Activity state & Flux$_{V}$ & Flux$_{B}$ & Flux$_{U}$ & Flux$_{W1}$ & Flux$_{M2}$ & Flux$_{W2}$ \\
 \hline
 GeV flare    & 1.14~$\pm$~0.22 & 1.23~$\pm$~0.19 & 1.08~$\pm$~0.10 & 1.06~$\pm$~0.10 & 1.15~$\pm$~0.09 & 0.99~$\pm$~0.08 \\
 Low activity & 0.81~$\pm$~0.17 & 0.57~$\pm$~0.09 & 0.46~$\pm$~0.05 & 0.56~$\pm$~0.04 & 0.62~$\pm$~0.05 & 0.66~$\pm$~0.04 \\
 \hline
\end{tabular}
}
\end{center}
\end{table}

\begin{table*}
{\small
\begin{center}
\caption{Summary of the parameters used/derived from the SED modeling. We assume the viewing angle as 3$^{\circ}$ and the characteristic temperature of the dusty torus as 600 K. For a black hole mass of 4.5 $\times$ 10$^7$ $M_{\odot}$ and a disk luminosity of 6 $\times$ 10$^{44}$ \lum, the size of the BLR and the dusty torus are 0.02 parsec and 0.5 parsec, respectively.}\label{tab:sed_par}
\begin{tabular}{lcc}
\tableline
\tableline
SED parameter                                                   & GeV flare   & Low activity \\
\tableline
Slope of particle spectral index before break energy ($p$)      & 1.45         & 2.3 \\
Slope of particle spectral index after break energy ($q$)       & 4.3          & 4.5 \\
Magnetic field in Gauss ($B$)                                   & 0.4          & 0.35\\
Particle energy density in erg cm$^{-3}$ ($U'_{\rm e}$)         & 0.01         & 0.01\\
Bulk Lorentz factor ($\Gamma$)                                  & 25           & 12\\
Doppler factor ($\delta$)                                       & 18           & 17 \\
Minimum Lorentz factor ($\gamma'_{\rm min}$)                     & 1            & 50\\
Break Lorentz factor ($\gamma'_{\rm b}$)                         & 1168         & 1584\\
Maximum Lorentz factor ($\gamma'_{\rm max}$)                     & 1.5e4        & 1.5e4\\
Dissipation distance in parsec ($R_{\rm diss}$)                 & 0.16         & 0.16\\
\hline
Jet power in electrons, in log scale ($P_{\rm e}$)               & 44.40      & 44.01\\
Jet power in magnetic field, in log scale ($P_{\rm B}$)          & 44.33      & 43.58\\
Radiative jet power, in log scale ($P_{\rm r}$)                  & 45.55      & 43.83\\
Jet power in protons, in log scale ($P_{\rm p}$)                 & 46.03      & 45.13\\
\tableline
\end{tabular}
\end{center}
}
\end{table*}

\begin{table}
\begin{center}
{
\small
\caption{Flux values for GeV flaring SEDs of $\gamma$-NLSy1 galaxies 1H 0323+342, PMN J0948+0022, and FSRQ 3C 279. The flux units are same as in Table \ref{tab:sed_flux1}.}\label{tab:sed_flux2}
\begin{tabular}{lcccccc}
\hline\hline
 & & & {\it Fermi}-LAT & &  & \\
 Name & & Flux$_{0.1-300~{\rm GeV}}$ & $\Gamma_{0.1-300~{\rm GeV}}$ & Test Statistic &\\
 \hline
 1H 0323+342    & & 10.00 $\pm$ 1.09 & 2.47 $\pm$ 0.11 & 284.22 & & \\
 PMN J0948+0022 & & 9.06 $\pm$ 0.87  & 2.65 $\pm$ 0.11 & 347.15 & & \\
 3C 279         & & 245.00 $\pm$ 4.85 & 2.05 $\pm$ 0.02 & 22673.92 & & \\
  \hline
 &  &  & {\it Swift}-XRT  & &  & \\
 Name & & Flux$_{0.3-10~{\rm keV}}$ & $\Gamma_{0.3-10~{\rm keV}}$ & C-Statistics & & \\
\hline
 1H 0323+342    & & 323.5$^{+18.30}_{-18.30}$ & 1.55$^{+0.08}_{-0.08}$ & 55.56/51 & & \\
 PMN J0948+0022 & & 85.89$^{+8.74}_{-7.59}$   & 1.55$^{+0.11}_{-0.11}$ & 21.03/27 & & \\
 3C 279         & & 661.4$^{+40.37}_{-41.27}$ & 1.25$^{+0.06}_{-0.06}$ & 53.44/69 & & \\
 \hline
 & & & {\it Swift}-UVOT & & &  \\
 Name & Flux$_{V}$ & Flux$_{B}$ & Flux$_{U}$ & Flux$_{W1}$ & Flux$_{M2}$ & Flux$_{W2}$ \\
 \hline
 1H 0323+342    & 21.50~$\pm$~0.75 & 21.70~$\pm$~0.82 & 25.10~$\pm$~1.07 & 21.60~$\pm$~1.27 & 24.60~$\pm$~1.41 & 24.30~$\pm$~1.28 \\
 PMN J0948+0022 & 5.90~$\pm$~0.44  & 5.46~$\pm$~0.30  & 4.95~$\pm$~0.23  & 3.97~$\pm$~0.19  & 5.75~$\pm$~0.39  & 5.02~$\pm$~0.26 \\
 3C 279         & 20.36~$\pm$~0.13$^*$ & --- & --- & --- & --- & 9.22~$\pm$~0.33 \\
 \hline
\end{tabular}
\tablecomments{$^*V$ band observations of 3C 279 were taken from Steward observatory.}
}
\end{center}
\end{table}

\newpage
\newpage

\newpage

\newpage

\end{document}